\documentclass[acmsmall,nonacm,screen]{acmart}

\usepackage{amsmath}

\usepackage{amssymb}
\usepackage{listings}
\usepackage{xcolor}
\usepackage{booktabs}
\usepackage{syntax}

\lstdefinelanguage{AutoLALA}{
  morekeywords={params,array,for,in,step,if,else,read,write,update},
  morecomment=[l]{//},
  sensitive=true,
  morestring=[b]",
  literate={..}{{\texttt{..}}}2,
}

\lstdefinelanguage{RustSnippet}{
  morekeywords={pub,struct,enum,fn,let,mut,impl,self,match,for,if,use,mod,crate,
                 Vec,String,HashMap,Option,Result,Box,Self,usize,i64,bool,
                 Clone,Debug,Serialize,Deserialize},
  morecomment=[l]{//},
  morecomment=[s]{/*}{*/},
  morestring=[b]",
  sensitive=true,
}

\lstdefinestyle{rust}{
  language=RustSnippet,
  basicstyle=\ttfamily\small,
  keywordstyle=\bfseries,
  commentstyle=\itshape\color{gray},
  frame=single,
  captionpos=b,
  xleftmargin=1.5em,
  framexleftmargin=1em,
  numbers=none,
}

\begin{document}

\title{AutoLALA: Automatic Loop Algebraic Locality Analysis\\for AI and HPC Kernels}

\author{Yifan Zhu}
\orcid{0009-0000-6718-7787}
\affiliation{
  \institution{University of Rochester}
  \city{Rochester}
  \state{NY}
  \country{USA}
}
\email{yifanzhu@rochester.edu}

\author{Yekai Pan}
\orcid{0009-0001-7639-1488}
\affiliation{
  \institution{University of Rochester}
  \city{Rochester}
  \state{NY}
  \country{USA}
}
\email{ypan34@u.rochester.edu}

\author{Yanghui Wu}
\orcid{0009-0005-2675-2811}
\affiliation{
  \institution{University of Rochester}
  \city{Rochester}
  \state{NY}
  \country{USA}
}
\email{ywu86@ur.rochester.edu}

\author{Chen Ding}
\orcid{0000-0003-4968-6659}
\affiliation{
  \institution{University of Rochester}
  \city{Rochester}
  \state{New York}
  \country{USA}
}

\begin{abstract}
Data movement is the main bottleneck in modern computing.
For loop-based programs common in high-performance computing (HPC) and AI workloads---including matrix multiplication, tensor contraction, stencil computation, and einsum operations---the cost of moving data through the memory hierarchy often exceeds the cost of arithmetic.
This paper presents \textsc{AutoLALA}, an open-source tool that takes a loop program written in a small domain-specific language (DSL), lowers it to polyhedral sets and maps, and produces closed-form symbolic formulas for reuse distance and data movement complexity.
The tool implements the fully symbolic locality analysis of Zhu et al.\ and the data movement distance (DMD) framework of Smith et al., computing reuse distance as the image of the access space under the access map---without stack simulation or Denning's recursive working-set formulation.
We describe the DSL syntax, its formal semantics, the polyhedral lowering that constructs timestamp spaces and access maps via affine transformations, and the chain of Barvinok counting operations that yields symbolic reuse-interval and reuse-distance distributions.
The system is implemented in Rust across three crates and integrates with the Barvinok library through safe bindings.
We provide a command-line tool and a web playground with LaTeX rendering of the output formulas.
The tool handles any affine loop nest, covering workloads such as tensor contractions, einsum expressions, stencil computations, and general polyhedral programs.
\end{abstract}

\maketitle

% --------------------------------------------------------------------------
\section{Introduction}
\label{sec:intro}

The gap between processor speed and memory latency has grown for decades.
Data movement---the traffic between a processor and its memory hierarchy---is now the main cost in both time and energy for many workloads~\cite{smith2022dmd,hong1981io}.
This is true for dense linear algebra in HPC, for tensor operations in machine learning, and for stencil computations in scientific simulation.

Time complexity counts operations while treating memory as free.
It cannot distinguish algorithms with the same operation count but different data movement, e.g., naive vs.\ tiled matrix multiplication, both $\Theta(N^3)$~\cite{smith2022dmd}.

\emph{Data Movement Distance} (DMD)~\cite{smith2022dmd} is a framework that assigns each memory access a cost proportional to the square root of its reuse distance---the number of distinct data elements accessed between two consecutive accesses to the same element.
The DMD of a program is the sum of these per-access costs.
This framework can distinguish algorithms with identical time complexity but different memory behavior.

Zhu et al.~\cite{zhu2026fullysymbolicanalysisloop} showed how to compute the reuse-distance distribution of an affine loop nest as a symbolic (quasi-)polynomial in the loop bounds.
Their method uses the polyhedral model: loop iterations form an integer set, memory accesses form an affine map, and reuse distances are obtained by composing these maps and counting integer points with the Barvinok library~\cite{barvinok2024,verdoolaege2007counting}.
The result is a set of closed-form formulas that hold for all input sizes and cache configurations, removing the need for simulation or profiling.

This paper describes \textsc{AutoLALA} (\textbf{Auto}matic \textbf{L}oop \textbf{A}lgebraic \textbf{L}ocality \textbf{A}nalysis), an open-source Rust implementation of this analysis pipeline.
The contributions are:

\begin{enumerate}
\item A \emph{domain-specific language} (DSL) for specifying affine loop programs with symbolic parameters, affine bounds, conditional guards, and typed memory accesses (\S\ref{sec:dsl}).
\item A \emph{polyhedral lowering} that translates DSL programs into timestamp spaces (integer sets) and access maps (affine relations), handling loop steps, statement interleaving, and conditional branches (\S\ref{sec:lowering}).
\item A \emph{symbolic locality analysis} that computes reuse-interval and reuse-distance distributions, assembles DMD formulas, and filters non-scaling branches---with reuse distance computed as the image of the access space under the access map (\S\ref{sec:analysis}).
\item A \emph{Rust implementation} organized as three crates---a core library, a CLI, and a web playground---with safe Barvinok bindings (\S\ref{sec:architecture}).
\end{enumerate}

The DSL accepts any affine loop nest.
This covers a broad class of HPC and AI workloads: dense matrix operations, tensor contractions and einsum expressions~\cite{blacher2024einsum,kjolstad2017taco}, stencil computations, image-processing pipelines~\cite{ragankelley2013halide}, and the PolyBench suite~\cite{pouchet2012polybench}.
Any loop program whose bounds and subscripts are affine functions of the enclosing iterators and symbolic parameters can be analyzed.

% --------------------------------------------------------------------------
\section{Background}
\label{sec:background}

\subsection{The Polyhedral Model}

The polyhedral model represents loop nests as integer polyhedra~\cite{aho2006compilers}.
A loop with affine bounds $\ell \le i < u$ defines a set of integer points $\{i \in \mathbb{Z} \mid \ell \le i < u\}$.
Nesting $d$ loops produces a $d$-dimensional polyhedron.
Memory accesses with affine subscripts (e.g., $A[i+1, 2j]$) define affine maps from the iteration space to a data space.
Tools such as ISL~\cite{verdoolaege2010isl} manipulate these sets and maps symbolically: union, intersection, composition, inversion, lexicographic ordering, and counting.

A \emph{quasi-polynomial} is a polynomial whose coefficients may depend on the residues of the variables modulo fixed constants.
The Barvinok library~\cite{barvinok2024,verdoolaege2007counting} computes the number of integer points in a parametric polytope as a quasi-polynomial in the parameters.
This is the key operation that turns polyhedral relations into closed-form symbolic formulas.

\subsection{Reuse Distance}

Reuse distance (also called LRU stack distance)~\cite{mattson1970evaluation,zhong2009program} is the number of distinct data elements accessed between two consecutive accesses to the same element.
Given an access trace $a_1, a_2, \ldots, a_n$, the reuse distance of access $a_t$ is the number of distinct elements in $\{a_{t'+1}, \ldots, a_{t-1}\}$ where $t'<t$ is the most recent prior access to the same element.
If $a_t$ is the first access to its element, the reuse distance is $\infty$ (a compulsory miss).

The reuse distance fully characterizes LRU cache behavior: an access with reuse distance $r$ hits in a fully associative LRU cache of size $c$ if and only if $r < c$~\cite{mattson1970evaluation}.

\subsection{Data Movement Distance}

The DMD framework~\cite{smith2022dmd} assigns each access a cost based on its reuse distance.
For a program with $n$ total accesses, of which $n_c$ are compulsory (first access to each element) and the remaining $n - n_c$ have finite reuse distances, the DMD is:
\begin{equation}
\label{eq:dmd}
\operatorname{DMD} = n_c + \sum_{\substack{t : \text{warm}\\\text{access}}} \sqrt{rd(t)}
\end{equation}
where $rd(t)$ is the reuse distance of access $t$.
The $\sqrt{\cdot}$ models the fact that data with smaller reuse distance is more likely to be cached, so its movement cost is lower.
Empirical studies have observed that cache miss rates scale as a power of reuse distance~\cite{hartstein2008sqrt2}, and the square root provides a robust approximation for modern hierarchies~\cite{smith2022dmd}.

When the reuse-distance distribution is given as a symbolic function---a set of regions $D_i$ with multiplicity $m_i$ and reuse distance $r_i$---the DMD becomes:
\begin{equation}
\label{eq:dmd-symbolic}
\operatorname{DMD} = n_c + \sum_{i} m_i \cdot \sqrt{r_i}
\end{equation}
where each $m_i$ and $r_i$ are quasi-polynomials in the symbolic parameters.

% --------------------------------------------------------------------------
\section{The AutoLALA DSL}
\label{sec:dsl}

\textsc{AutoLALA} accepts programs in a small DSL designed for specifying affine loop nests.
We describe its syntax formally and illustrate it with examples.

\subsection{Formal Syntax}

The grammar is given in BNF-like notation.
Terminals are in \texttt{typewriter} font; nonterminals are in $\langle$angle brackets$\rangle$.

\begin{grammar}
<program> ::= <param-decl>* <array-decl>* <stmt>*

<param-decl> ::= `params' <id-list> `;'

<array-decl> ::= `array' <id> `[' <expr-list> `]' `;'

<stmt> ::= <for-stmt> | <if-stmt> | <access-stmt>

<for-stmt> ::= `for' <id> `in' <expr> `..' <expr> <step-spec>? <block>

<step-spec> ::= `step' <integer>

<if-stmt> ::= `if' <cond-list> <block> (`else' <block>)?

<block> ::= `\{' <stmt>* `\}'

<access-stmt> ::= <access-kind> <id> `[' <expr-list> `]' `;'

<access-kind> ::= `read' | `write' | `update'

<cond-list> ::= <comparison> (`\&\&' <comparison>)*

<comparison> ::= <expr> <cmp-op> <expr>

<cmp-op> ::= `<' | `<=' | `==' | `>=' | `>'

<expr> ::= <expr> (`+' | `-') <mul-expr> | <mul-expr>

<mul-expr> ::= <mul-expr> (`*' | `/') <unary> | <unary>

<unary> ::= `-' <unary> | `(' <expr> `)' | <integer> | <id>
\end{grammar}

A \emph{program} consists of zero or more parameter declarations, zero or more array declarations, and a sequence of statements.
Parameters are symbolic constants (e.g., matrix dimensions); arrays are declared with their extents as affine expressions of the parameters.

\subsection{Affine Restriction}

All expressions in bounds, subscripts, conditions, and array extents must be \emph{affine}: degree at most~1 in any variable.
Concretely:
\begin{itemize}
\item Multiplication requires at least one operand to be a compile-time constant.
The product $i \cdot j$ where both $i$ and $j$ are loop variables is rejected.
\item Division (\texttt{/}) computes integer floor division and requires a constant divisor.
\item Variables in an expression must be symbolic parameters or enclosing loop iterators.
\end{itemize}

This restriction is necessary for the polyhedral model: the iteration domain must be an integer polyhedron and the access function must be an affine map.

\subsection{Access Kinds}

Each memory access is tagged as \texttt{read}, \texttt{write}, or \texttt{update}.
For the purpose of locality analysis, all three are treated uniformly: they contribute equally to the timestamp sequence and the access map.
The distinction is retained in the AST for potential future use in cost models that distinguish reads from writes.

\subsection{Example: Matrix Multiplication}

\begin{lstlisting}[caption={Matrix multiplication in the AutoLALA DSL.},label={lst:matmul}]
params M, N, K;
array A[M, K];
array B[K, N];
array C[M, N];

for i in 0 .. M {
  for j in 0 .. N {
    for k in 0 .. K {
      read C[i, j];
      read A[i, k];
      read B[k, j];
      write C[i, j];
    }
  }
}
\end{lstlisting}

This program specifies a triply-nested loop with four memory accesses per iteration.
The symbolic parameters $M$, $N$, $K$ are the matrix dimensions.
The tool produces symbolic formulas for the reuse-distance distribution and the DMD as functions of $M$, $N$, and $K$.

\subsection{Example: 1D Jacobi Stencil}

\begin{lstlisting}[caption={1D Jacobi stencil.},label={lst:stencil}]
params N, T;
array A[N];
array B[N];

for t in 0 .. T {
  for i in 1 .. N - 1 {
    read A[i - 1];
    read A[i];
    read A[i + 1];
    write B[i];
  }
  for i in 1 .. N - 1 {
    read B[i];
    write A[i];
  }
}
\end{lstlisting}

The stencil accesses \texttt{A[i-1]}, \texttt{A[i]}, and \texttt{A[i+1]}, all of which are affine.
The tool computes the reuse distances between consecutive timesteps and across the $i$-loop.

\subsection{Example: Einsum / Tensor Contraction}

Tensor contractions, frequently expressed as einsum operations in deep-learning frameworks, are naturally affine.
A contraction $C_{ij} = \sum_k A_{ik} B_{kj}$ is exactly the matrix multiplication of Listing~\ref{lst:matmul}.
Higher-order contractions such as $D_{ij} = \sum_{k,l} A_{ikl} B_{klj}$ add more loops but keep all subscripts affine.
Similarly, batched operations like $C_{bij} = \sum_k A_{bik} B_{bkj}$ simply add a batch loop.
All of these fall within the scope of \textsc{AutoLALA}.

% --------------------------------------------------------------------------
\section{Polyhedral Lowering}
\label{sec:lowering}

This section describes how the tool translates a DSL program into the polyhedral objects required for locality analysis: a \emph{timestamp space} (an integer set) and an \emph{access map} (an affine relation).

\subsection{Timestamp Space}
\label{sec:timestamp}

The timestamp space encodes the execution order of all memory accesses in the program.
Each access is assigned a point in a multi-dimensional integer space whose lexicographic order matches the execution order.

\paragraph{Loop dimensions.}
A $\texttt{for}$ loop with iterator $i$, lower bound $\ell$, upper bound $u$, and step $s$ contributes one dimension.
Rather than storing the iterator value directly, the tool stores the \emph{ordinal} iteration count $o_i \in \{0, 1, \ldots, \lceil(u - \ell)/s\rceil - 1\}$.
The actual iterator value is recovered as $i = o_i \cdot s + \ell$.
The ordinal representation has two advantages: the lower bound is always~0, and the step is always~1.

Formally, for a single loop, the contribution to the timestamp space is:
\begin{equation}
\label{eq:loop-dim}
\mathcal{T}_{\texttt{for}} = \bigl\{ o_i \in \mathbb{Z} \;\big|\; 0 \le o_i \;\wedge\; o_i < \lceil(u - \ell)/s\rceil \bigr\}
\end{equation}
For nested loops, the timestamp space is the Cartesian product constrained by the conjunction of all loop bounds:
\begin{equation}
\label{eq:timestamp}
\mathcal{T} = \bigl\{ (o_1, \ldots, o_d) \in \mathbb{Z}^d \;\big|\; \bigwedge_{k=1}^{d} 0 \le o_k < \lceil(u_k - \ell_k)/s_k\rceil \bigr\}
\end{equation}
where the bounds $\ell_k$ and $u_k$ may depend on enclosing loop ordinals and symbolic parameters.

\paragraph{Statement interleaving.}
When a block contains multiple statements, each statement $j$ (numbered from~0) gets a dedicated ``selector'' dimension set to the constant~$j$.
For a block with $m$ statements $S_0, \ldots, S_{m-1}$, the timestamp space is:
\begin{equation}
\mathcal{T}_{\text{block}} = \bigcup_{j=0}^{m-1} \bigl\{ (j) \times \mathcal{T}_{S_j} \bigr\}
\end{equation}
where $\mathcal{T}_{S_j}$ is the timestamp space of statement $S_j$, padded with zero-constrained dimensions so that all branches have the same dimensionality.
The selector dimension is placed at the depth of the enclosing block, and all sub-statements share the same parameter space.

\paragraph{Conditional branches.}
An \texttt{if} statement with condition $\phi$ is handled by intersecting the then-branch timestamp space with the condition set and the else-branch with its complement:
\begin{equation}
\mathcal{T}_{\texttt{if}} = (\mathcal{T}_{\text{then}} \cap \llbracket\phi\rrbracket) \;\cup\; (\mathcal{T}_{\text{else}} \cap \llbracket\neg\phi\rrbracket)
\end{equation}
The condition $\phi$ is a conjunction of affine comparisons, so $\llbracket\phi\rrbracket$ is a polyhedron.

\subsection{Access Map}
\label{sec:access-map}

The access map $\mathcal{A}: \mathcal{T} \to \mathcal{D}$ sends each timestamp to the memory location it accesses.
The data space $\mathcal{D}$ is a tuple:
\begin{equation}
\label{eq:data-space}
\mathcal{D} = (\textit{array\_id},\; d_1, \ldots, d_r,\; [b,\; s])
\end{equation}
where $\textit{array\_id}$ is an integer identifying the array, $d_1, \ldots, d_r$ are the array subscript values padded to a uniform rank $r = \max_{\text{arrays}} \operatorname{rank}$, and the optional components $b$ and $s$ handle cache blocking and set-associativity.

\paragraph{Affine subscript transformation.}
For an access $\texttt{read A}[f_1(\vec{i},\vec{p}), \ldots, f_k(\vec{i},\vec{p})]$, each subscript $f_j$ is an affine function of the enclosing loop iterators $\vec{i}$ and the symbolic parameters $\vec{p}$.
Since the iterators are stored in ordinal form, the tool converts back: if the $q$-th loop has lower bound $\ell_q$ and step $s_q$, and the ordinal dimension is $o_q$, then a reference to iterator $i_q$ in the subscript is replaced by $o_q \cdot s_q + \ell_q$.
The resulting map component is:
\begin{equation}
f_j(\vec{i},\vec{p}) = f_j(o_1 \cdot s_1 + \ell_1, \ldots, o_d \cdot s_d + \ell_d, \vec{p})
\end{equation}
This is still affine in the ordinals $o_q$ and parameters $\vec{p}$, so the access map is an affine relation.

\paragraph{Cache blocking.}
When the tool is invoked with a block size $B > 1$, the last subscript dimension is replaced by $\lfloor d_r / B \rfloor$.
This models the effect of cache-line granularity: two accesses to the same cache line map to the same data-space point.

\paragraph{Set-associative tagging.}
When the number of sets $S > 1$, an extra dimension $d_r \bmod S$ is appended.
This models set-associative caches where the set index determines which cache set an address maps to.

\paragraph{Construction.}
The access map for each statement is built by constructing an affine function list:
\begin{equation}
\mathcal{A}_{\text{stmt}} = \bigl\{ \vec{o} \to (\textit{id},\; f_1(\vec{o}, \vec{p}),\; \ldots,\; f_r(\vec{o}, \vec{p})) \bigr\}
\end{equation}
Block-level maps are formed by unioning the maps of all sub-statements (after alignment, analogous to timestamp space construction).
The final access map is intersected with the timestamp space to restrict it to reachable points:
\begin{equation}
\mathcal{A} = \mathcal{A}_{\text{raw}} \cap (\mathcal{T} \times \mathcal{D})
\end{equation}

% --------------------------------------------------------------------------
\section{Symbolic Locality Analysis}
\label{sec:analysis}

Given the timestamp space $\mathcal{T}$ and the access map $\mathcal{A}: \mathcal{T} \to \mathcal{D}$, the tool computes the reuse-interval (RI) distribution, the reuse-distance (RD) distribution, and the DMD formula.
All intermediate objects are polyhedral relations manipulated symbolically.

\subsection{Lexicographic Ordering}

The execution order is the lexicographic order on $\mathcal{T}$.
The tool constructs two relations:
\begin{align}
\prec &= \{(\vec{t}, \vec{t}') \in \mathcal{T}^2 \mid \vec{t} <_{\text{lex}} \vec{t}'\} \\
\preceq &= \{(\vec{t}, \vec{t}') \in \mathcal{T}^2 \mid \vec{t} \le_{\text{lex}} \vec{t}'\}
\end{align}
These are standard polyhedral operations provided by ISL~\cite{verdoolaege2010isl} as \texttt{lex\_lt} and \texttt{lex\_le}, restricted to the timestamp space.

\subsection{Same-Element Relation}

Two timestamps access the same data element if they map to the same point in $\mathcal{D}$:
\begin{equation}
\label{eq:same-elem}
\mathcal{S} = \mathcal{A} \circ \mathcal{A}^{-1} = \{(\vec{t}, \vec{t}') \mid \mathcal{A}(\vec{t}) = \mathcal{A}(\vec{t}')\}
\end{equation}
This is the composition of the access map with its inverse.
Two timestamps $(\vec{t}, \vec{t}')$ are in $\mathcal{S}$ if and only if they access the same array element (or the same cache line, when blocking is enabled).

\subsection{Immediate Predecessor}

Among all same-element pairs where $\vec{t}'$ comes before $\vec{t}$, the \emph{immediate predecessor} of $\vec{t}$ is the latest such $\vec{t}'$:
\begin{equation}
\label{eq:imm-next}
\mathcal{N} = \operatorname{lexmin}(\mathcal{S} \cap \prec)
\end{equation}
Here $\operatorname{lexmin}$ selects the lexicographically smallest image for each domain element, giving the \emph{immediate next} same-element access.
Its inverse $\mathcal{P} = \mathcal{N}^{-1}$ gives the immediate predecessor.

\subsection{Reuse Interval}

The reuse interval of a warm access $\vec{t}$ is the set of timestamps between its immediate predecessor and itself.
The tool computes this as:
\begin{equation}
\label{eq:ri-relation}
\mathcal{I} = \mathcal{P} \circ \prec
\end{equation}
This maps each warm-access timestamp $\vec{t}$ to all timestamps strictly after its predecessor and (by further intersection) up to and including $\vec{t}$:
\begin{equation}
\label{eq:ri}
\mathcal{RI} = \mathcal{I} \cap \succeq
\end{equation}
where $\succeq$ is $\preceq^{-1}$.
For a given $\vec{t}$, the set $\mathcal{RI}(\vec{t}) = \{\vec{t}' \mid \mathcal{P}(\vec{t}) \prec \vec{t}' \preceq \vec{t}\}$ contains all timestamps in the reuse window.

The \emph{reuse-interval value} (the number of accesses in the window) is the cardinality:
\begin{equation}
ri(\vec{t}) = |\mathcal{RI}(\vec{t})|
\end{equation}
Applying the Barvinok cardinality operation to the relation $\mathcal{RI}$ yields $ri$ as a piecewise quasi-polynomial in the loop iterators and symbolic parameters.

\subsection{Reuse Distance as Image under Access Map}
\label{sec:rd-image}

The reuse distance counts the number of \emph{distinct data elements} in the reuse window.
Traditional approaches compute this via stack simulation or the inclusion-exclusion counting of Denning's working-set recursion~\cite{denning2021workingset}.
\textsc{AutoLALA} takes a different route, following the approach of Zhu et al.~\cite{zhu2026fullysymbolicanalysisloop}: it computes the reuse distance as the cardinality of the \emph{image} of the reuse-interval window under the access map.

Concretely, the tool composes the reuse-interval relation with the access map:
\begin{equation}
\label{eq:rd-relation}
\mathcal{RD} = \mathcal{RI} \circ \mathcal{A}
\end{equation}
For a given warm-access timestamp $\vec{t}$, the set $\mathcal{RD}(\vec{t}) = \mathcal{A}(\mathcal{RI}(\vec{t}))$ is the set of distinct data elements accessed in the reuse window.
The reuse distance is:
\begin{equation}
\label{eq:rd-value}
rd(\vec{t}) = |\mathcal{RD}(\vec{t})|
\end{equation}
Applying Barvinok's cardinality operation to $\mathcal{RD}$ yields $rd$ as a piecewise quasi-polynomial.

This formulation avoids any recursive working-set computation.
The reuse distance is obtained in a single composition followed by a cardinality count.
The composition $\mathcal{RI} \circ \mathcal{A}$ is a standard polyhedral operation: for each timestamp $\vec{t}$, it collects all data elements $\mathcal{A}(\vec{t}')$ for $\vec{t}'$ in the reuse window, then the cardinality counts the number of distinct such elements.

\subsection{Distribution Processing}

The piecewise quasi-polynomials for RI and RD are organized by \emph{value}: pieces with the same quasi-polynomial expression for $ri(\vec{t})$ (or $rd(\vec{t})$) are grouped.
Within each group, the domain (the set of timestamps with that reuse distance) is reported as a union of parametric polyhedra, and its cardinality gives the \emph{multiplicity}---the number of accesses with that reuse distance.

For each value group~$i$, the tool extracts:
\begin{itemize}
\item The reuse-distance value $r_i$: a quasi-polynomial in the parameters.
\item A set of regions $D_{i,1}, D_{i,2}, \ldots$, each a parametric polyhedron.
\item The multiplicity $m_{i,j} = |D_{i,j}|$: the number of timestamps in each region.
\end{itemize}

\subsection{DMD Assembly}

The tool computes three access counts:
\begin{align}
n_{\text{total}} &= |\mathcal{T}| \\
n_{\text{warm}} &= \sum_{i,j} m_{i,j} \\
n_c &= n_{\text{total}} - n_{\text{warm}}
\end{align}
where $n_c$ is the number of compulsory (cold) accesses.
The DMD formula is then:
\begin{equation}
\label{eq:dmd-assembled}
\operatorname{DMD} = n_c + \sum_{i,j} m_{i,j} \cdot \sqrt{r_i}
\end{equation}
Each $m_{i,j}$ and $r_i$ is a quasi-polynomial in the symbolic parameters, so the entire DMD formula is symbolic.

\subsection{Region Scaling Filter}
\label{sec:filter}

Not all regions in the reuse-distance distribution contribute to the asymptotic DMD.
Some regions are ``pinned'': they exist only for specific constant values of a parameter (e.g., $K = 2$) or are bounded from above (e.g., $K \le 512$).
These do not scale with the problem size and are irrelevant for asymptotic analysis.

The tool filters a region $D$ based on its constraints:
\begin{itemize}
\item If any named parameter dimension is fixed by an equality constraint (e.g., $N = 5$), the region does not scale.
\item If any named parameter dimension has an upper-bound constraint (e.g., $N \le c$), the region does not scale.
\end{itemize}
Only regions that pass this filter contribute to the DMD formula.
The full RI and RD distributions are still reported for diagnostic purposes.

% --------------------------------------------------------------------------
\section{System Architecture}
\label{sec:architecture}

\textsc{AutoLALA} is implemented in Rust (edition 2024, minimum version 1.92) as a Cargo workspace with three crates.

\subsection{Crate Organization}

\paragraph{\texttt{dmd-core}.}
The core library implements the full analysis pipeline.
It contains six modules:
\begin{itemize}
\item \texttt{lexer}: tokenization using the \texttt{logos} crate.
Tokens include keywords (\texttt{params}, \texttt{array}, \texttt{for}, \texttt{in}, \texttt{step}, \texttt{if}, \texttt{else}, \texttt{read}, \texttt{write}, \texttt{update}), operators (\texttt{+}, \texttt{-}, \texttt{*}, \texttt{/}, \texttt{\&\&}, comparisons), delimiters, identifiers, and integer literals.
\item \texttt{grammar} / \texttt{parser}: an LALRPOP grammar that produces an AST.
LALRPOP generates an LR(1) parser from the grammar specification at build time.
\item \texttt{ast}: type definitions for the abstract syntax tree (programs, loops, conditionals, accesses, affine expressions).
\item \texttt{semantics}: validates the AST.
Checks include: unique parameter names, correct array ranks, no loop-variable shadowing, affine expressions only (rejects variable$\times$variable products), and positive loop steps.
\item \texttt{polyhedral}: the lowering and analysis engine.
Constructs timestamp spaces and access maps, computes RI/RD distributions and DMD formulas.
All Barvinok calls are serialized under a global mutex because the library's context initialization is not thread-safe.
\item \texttt{formula}: algebraic expression type with simplification (flattening, coefficient collection, constant folding, perfect-square extraction) and dual rendering to plain text and LaTeX.
\end{itemize}

\paragraph{\texttt{dmd-cli}.}
A command-line tool built with \texttt{clap}.
It reads a DSL file (or stdin), invokes \texttt{dmd-core}, and prints the analysis report as plain text or JSON.
Options include \texttt{-{}-block-size}, \texttt{-{}-num-sets}, \texttt{-{}-max-operations}, and \texttt{-{}-approximation-method}.

\paragraph{\texttt{dmd-playground}.}
A web application using \texttt{axum} (async Rust web framework) and \texttt{tokio}.
The frontend uses the Monaco editor for DSL editing with custom syntax highlighting and KaTeX for LaTeX formula rendering.
Analysis jobs are submitted via \texttt{POST /api/tasks} and polled via \texttt{GET /api/tasks/\{id\}}.
Concurrency is bounded by a tokio semaphore (default: 4), and each job has a timeout (default: 30 seconds).

\subsection{Barvinok Integration}

The tool uses \texttt{barvinok-rs}, a Rust binding to the Barvinok/ISL C libraries.
The binding's build script (\texttt{barvinok-sys}) bundles the C sources for Barvinok, ISL, and PolyLib and compiles them via autotools.
Host dependencies are GMP, NTL, and libclang.

Barvinok is invoked with the \texttt{-{}-approximation-method=scale} flag, which uses the scale approximation for counting quasi-polynomials.
All Barvinok operations (context creation, set/map construction, cardinality computation) happen inside a scoped context that bounds the maximum number of operations to prevent runaway computations on programs with high-dimensional iteration spaces.

\subsection{Formula Rendering}

The \texttt{FormulaExpr} type represents symbolic expressions:
\begin{lstlisting}[style=rust,caption={Core formula expression type.}]
pub enum FormulaExpr {
    Rational { numerator: i64, denominator: i64 },
    Symbol(String),
    Raw { plain: String, latex: String },
    Add(Vec<FormulaExpr>),
    Mul(Vec<FormulaExpr>),
    Div(Box<FormulaExpr>, Box<FormulaExpr>),
    Pow(Box<FormulaExpr>, u32),
    Sqrt(Box<FormulaExpr>),
}
\end{lstlisting}
Each variant renders to both plain text (e.g., \texttt{sqrt(N + M) * K}) and LaTeX (e.g., $\sqrt{N + M} \cdot K$).
The \texttt{FormulaFormatter} converts Barvinok quasi-polynomials term-by-term into \texttt{FormulaExpr} values, which are then simplified.

\subsection{Pipeline Summary}

The full pipeline is: DSL source $\to$ lexer (logos) $\to$ parser (LALRPOP) $\to$ AST $\to$ semantic validation $\to$ timestamp space $\mathcal{T}$ + access map $\mathcal{A}$ (ISL) $\to$ reuse analysis ($\mathcal{RI}$, $\mathcal{RD}$) $\to$ Barvinok counting $\to$ distribution processing $\to$ DMD assembly + region filter $\to$ formula rendering (plain text + LaTeX).

% --------------------------------------------------------------------------
\section{Walkthrough}
\label{sec:walkthrough}

We trace the analysis of a program to show how polyhedral objects are built.
Consider:
\begin{lstlisting}[caption={Nested loop with reuse.},label={lst:simple}]
params N, M;
array A[N, M];
array B[M];

for i in 0 .. N {
  for j in 0 .. M {
    read A[i, j];
    read B[j];
  }
}
\end{lstlisting}
Array $B[j]$ is reused across outer iterations: within a fixed $i$, consecutive $j$-iterations access distinct elements of $B$, but across $i$-iterations the same $B[j]$ is reused.

\paragraph{Timestamp space.}
The loop nest has depth~2 ($i$, $j$) with two statements in the body, giving a selector $t_0 \in \{0, 1\}$:
\[
\mathcal{T} = \{(i, j, t_0) \mid 0 \le i < N \;\wedge\; 0 \le j < M \;\wedge\; 0 \le t_0 \le 1\}
\]

\paragraph{Access map.}
Let $A$ have ID 0, $B$ have ID 1.
The maximum rank is 2, so $B$'s subscript is padded:
$\mathcal{A}_A = \{(i, j, 0) \to (0, i, j)\}$,
$\mathcal{A}_B = \{(i, j, 1) \to (1, 0, j)\}$.

\paragraph{Reuse analysis.}
The same-element relation pairs timestamps accessing the same element.
For $B[j]$: every $(i, j, 1)$ for any $i$ accesses it.
The immediate predecessor of $(i, j, 1)$ (when $i > 0$) is $(i{-}1, j, 1)$, with a reuse interval spanning ${\sim}2M$ accesses.
The reuse distance counts distinct elements in that window; since $A$ and $B$ have different IDs, the tool computes this symbolically as a quasi-polynomial in $N$ and $M$.

% --------------------------------------------------------------------------
\section{Scope of Applicability}
\label{sec:scope}

The analysis applies to any loop nest where all bounds and subscripts are affine in the enclosing iterators and symbolic parameters.
This covers several important classes of HPC and AI workloads:

\paragraph{Dense linear algebra.}
Matrix multiplication, LU, Cholesky, QR, and other routines are affine.
The PolyBench/C suite~\cite{pouchet2012polybench} contains 30 such benchmarks.

\paragraph{Tensor contractions and einsum.}
A contraction $C_{i_1 \ldots i_m} = \sum_{k_1 \ldots k_n} A_{i_1 \ldots i_m k_1 \ldots k_n} B_{k_1 \ldots k_n j_1 \ldots j_p}$ is a multi-loop nest with affine subscripts.
Einsum expressions~\cite{blacher2024einsum} in PyTorch, NumPy, and TensorFlow map directly to contractions.
TACO~\cite{kjolstad2017taco} generates code of this form.
Prior analytical cache models for tensor contractions~\cite{li2019analytical,cociorva2002spacetime} targeted specific patterns; \textsc{AutoLALA} handles the general case.

\paragraph{Stencils and convolutions.}
Jacobi, Gauss-Seidel, and higher-dimensional stencils have affine access patterns, as do convolution layers expressed as nested loops.
Image-processing pipelines~\cite{ragankelley2013halide} use stencils extensively.

\paragraph{Limitations.}
Data-dependent control flow, indirect access ($A[B[i]]$), and non-affine subscripts ($A[i \cdot j]$) are outside the scope of the polyhedral model.

% --------------------------------------------------------------------------
\section{Related Work}
\label{sec:related}

\paragraph{Cache miss equations.}
Ghosh et al.~\cite{ghosh1999cache} derived symbolic cache miss counts via Diophantine equations.
Bao et al.~\cite{bao2018analytical} built a polyhedral analytical cache model.
Pitchanathan and Grosser~\cite{pitchanathan2024falcon} presented Falcon, a scalable analytical model.
These target specific cache configurations; \textsc{AutoLALA} produces cache-independent reuse-distance distributions.

\paragraph{Reuse distance analysis.}
Mattson et al.~\cite{mattson1970evaluation} introduced stack-distance analysis.
Zhong et al.~\cite{zhong2009program} developed compiler-based reuse-distance analysis; Beyls and D'Hollander~\cite{beyls2005generating} applied it to cache hints.
The relational theory of locality~\cite{yuan2019relational} unified reuse distance with higher-order theories~\cite{xiang2013hotl}.
\textsc{AutoLALA} builds on Zhu et al.~\cite{zhu2026fullysymbolicanalysisloop}, computing reuse distance as parametric quasi-polynomials.

\paragraph{Data movement complexity.}
Hong and Kung~\cite{hong1981io} introduced I/O complexity.
Olivry et al.~\cite{olivry2020lower} derived parametric lower bounds for affine programs.
Smith et al.~\cite{smith2022dmd} proposed the DMD framework that \textsc{AutoLALA} implements.

\paragraph{Polyhedral tools.}
ISL~\cite{verdoolaege2010isl} and Barvinok~\cite{barvinok2024,verdoolaege2007counting} provide the underlying operations.
MLIR~\cite{lattner2021mlir} and Polygeist~\cite{moses2021polygeist} provide polyhedral compiler infrastructure.
\textsc{AutoLALA} replaces an earlier MLIR-based path with a standalone DSL.

\paragraph{Parallel locality.}
Liu et al.~\cite{liu2024pluss} extended symbolic locality analysis to parallel programs with a symbolic thread count.
Extending \textsc{AutoLALA} to parallel loops using their techniques is a natural direction for future work.

% --------------------------------------------------------------------------
\section{Usage}
\label{sec:usage}

\subsection{Command-Line Interface}

The CLI reads a DSL file and prints the DMD formula, access counts, and RI/RD distributions:
\begin{verbatim}
$ cargo run -p dmd-cli -- --input matmul.dsl
\end{verbatim}
Flags include \texttt{-{}-json} for structured output, \texttt{-{}-block-size $B$} for cache-line modeling, and \texttt{-{}-num-sets $S$} for set-associative modeling.
Input is read from a file (\texttt{-{}-input}) or from stdin.

\subsection{Web Playground}

Running \texttt{cargo run -p dmd-playground -{}- -{}-port 3000} starts an interactive web interface with a Monaco-based DSL editor (with syntax highlighting), server-side analysis, and KaTeX-rendered output formulas.
The backend limits concurrency and enforces per-job timeouts.

% --------------------------------------------------------------------------
\section{Conclusion}
\label{sec:conclusion}

We presented \textsc{AutoLALA}, a tool for automatic symbolic locality analysis of affine loop programs.
Given a program in a small DSL, the tool lowers it to polyhedral sets and maps, computes reuse-interval and reuse-distance distributions using Barvinok's integer-point counting, and assembles a closed-form DMD formula.
The reuse distance is computed as the image of the access space under the access map, avoiding stack simulation.
The result is a set of symbolic quasi-polynomials that characterize the data movement behavior of the program for all input sizes.

The tool covers a broad range of HPC and AI workloads---matrix operations, tensor contractions, einsum expressions, stencil computations---and is implemented in Rust with a CLI and a web playground.
Future directions include extending the analysis to parallel programs following Liu et al.~\cite{liu2024pluss}, supporting tiling transformations, and integrating with compiler frameworks such as MLIR~\cite{lattner2021mlir} to analyze programs extracted from production code.

\bibliographystyle{ACM-Reference-Format}
\bibliography{references}

\end{document}